\documentclass[twocolumn,showpacs,showkeys,preprintnumbers,amsmath,amssymb,prc,superscriptaddress]{revtex4-1}

\usepackage{sistyle}
\usepackage[version=3]{mhchem}
\usepackage{xspace}
\usepackage{graphicx}
\usepackage{multirow}
\usepackage{setspace}
\usepackage[usenames,dvipsnames]{color}
\usepackage{diagbox}

\usepackage{booktabs}		

\newcommand{\tref}[1]{Table~\ref{#1}}
\newcommand{\fref}[1]{Fig.~\ref{#1}}
\newcommand{\sref}[1]{Sec.~\ref{#1}}

\newcommand\Artt{\ce{^{33}Ar}\xspace}
\newcommand\Cltt{\ce{^{33}Cl}\xspace}
\newcommand{\Stt}{\ce{^{32}S}\xspace}
\newcommand\Ar{\ce{^{31}Ar}\xspace}
\newcommand\Si{\ce{^{28}Si}\xspace}
\renewcommand{\S}{\ce{^{30}S}\xspace}
\newcommand\Cl{\ce{^{31}Cl}\xspace}
\renewcommand\P{\ce{^{29}P}\xspace}
\newcommand\Qep{$Q_{3p}$\xspace}
\newcommand\Qtp{$Q_{2p}$\xspace}
\newcommand\SIU[1]{\,\SI{}{#1}}
\newcommand\SIE[3]{\num{#1}(\num{#2})\,\SI{}{#3}}

\newcommand\rec{$\ce{^{29}P}(p,\gamma)\ce{^{30}S}$\xspace}

\begin{document}
\title{Multi-particle emission in the decay of \Ar.}

\author{G. T. Koldste}
\affiliation{Department of Physics and Astronomy, Aarhus University, DK-8000 Aarhus C, Denmark}
\author{B. Blank}
\affiliation{Centre d'{\'E}tudes Nucl{\'e}aire de Bordeaux-Gradignan, CNRS/IN2P3 -- Universit{\'e} Bordeaux I, F-33175 Gradignan Cedex, France}
\author{M. J. G. Borge}
\affiliation{Instituto de Estructura de la Materia, CSIC, E-28006 Madrid, Spain}
\author{J. A. Briz}
\affiliation{Instituto de Estructura de la Materia, CSIC, E-28006 Madrid, Spain}
\author{M. \surname{Carmona-Gallardo}}
\affiliation{Instituto de Estructura de la Materia, CSIC, E-28006 Madrid, Spain}
\author{L. M. Fraile} 
\affiliation{Grupo de F{\'i}sica Nuclear, Universidad Complutense, E-28040 Madrid, Spain}
\author{H. O. U. Fynbo}
\affiliation{Department of Physics and Astronomy, Aarhus University, DK-8000 Aarhus C, Denmark}
\author{J. Giovinazzo}
\affiliation{Centre d'{\'E}tudes Nucl{\'e}aire de Bordeaux-Gradignan, CNRS/IN2P3 -- Universit{\'e} Bordeaux I, F-33175 Gradignan Cedex, France}
\author{B. D. Grann}
\affiliation{Department of Physics and Astronomy, Aarhus University, DK-8000 Aarhus C, Denmark}
\author{J. G. Johansen}
\altaffiliation[Present address: ]{Institut f{\"u}r Kernphysik, Technische Universit{\"a}t Darmstadt, D-64289 Darmstadt, Germany}
\affiliation{Department of Physics and Astronomy, Aarhus University, DK-8000 Aarhus C, Denmark}
\author{A. Jokinen}
\affiliation{Department of Physics, University of Jyv{\"a}skyl{\"a}, FIN-40351 Jyv{\"a}skyl{\"a}, Finland}
\author{B. Jonson}
\affiliation{Fundamental Fysik, Chalmers Tekniska H{\"o}gskola, S-41296 G{\"o}teborg, Sweden}
\author{T. \surname{Kurturkian-Nieto}}
\affiliation{Centre d'{\'E}tudes Nucl{\'e}aire de Bordeaux-Gradignan, CNRS/IN2P3 -- Universit{\'e} Bordeaux I, F-33175 Gradignan Cedex, France}
\author{J. H. Kusk}
\affiliation{Department of Physics and Astronomy, Aarhus University, DK-8000 Aarhus C, Denmark}
\author{T. Nilsson}
\affiliation{Fundamental Fysik, Chalmers Tekniska H{\"o}gskola, S-41296 G{\"o}teborg, Sweden}
\author{A. Perea} 
\affiliation{Instituto de Estructura de la Materia, CSIC, E-28006 Madrid, Spain}
\author{V. Pesudo}
\affiliation{Instituto de Estructura de la Materia, CSIC, E-28006 Madrid, Spain}
\author{E. Picado}
\affiliation{Grupo de F{\'i}sica Nuclear, Universidad Complutense, E-28040 Madrid, Spain}
\affiliation{Secci{\'o}n de Radiaciones, Universidad Nacional, Heredia, Costa Rica}
\author{K. Riisager}
\affiliation{Department of Physics and Astronomy, Aarhus University, DK-8000 Aarhus C, Denmark}
\author{A. Saastamoinen}
\altaffiliation[Present address: ]{Cyclotron Institute, Texas A\&M University, College Station, TX 77843-3366, USA}
\affiliation{Department of Physics, University of Jyv{\"a}skyl{\"a}, FIN-40351 Jyv{\"a}skyl{\"a}, Finland}
\author{O. Tengblad}
\affiliation{Instituto de Estructura de la Materia, CSIC, E-28006 Madrid, Spain}
\author{J.-C. Thomas}
\affiliation{GANIL, CEA/DSM-CNRS/IN2P3, F-14076 Caen Cedex 5, France}
\author{J. \surname{Van de Walle}}
\affiliation{CERN, CH-1211 Geneva 23, Switzerland}

\date{\today}

\begin{abstract}
A multi-hit capacity setup was used to study the decay of the dripline nucleus \Ar, produced at the ISOLDE facility at CERN.

A spectroscopic analysis of the $\beta$-delayed three-proton decay of \Ar is presented for the first time together with a quantitative analysis of the $\beta$-delayed 2p$\gamma$-decay. A new method for determination of the spin of low-lying levels in the $\beta$p-daughter \S using proton-proton angular correlations is presented and used for the level at $\SI{5.2}{MeV}$, which is found to be either a $3^+$ or $4^+$ level, with the data pointing towards the $3^+$. The half-life of \Ar is found to be $\SIE{15.1}{3}{ms}$. An improved analysis of the Fermi $\beta$-strength gives a total measured branching for the $\beta$3p-decay of $\SIE{3.60}{44}{\%}$, which is lower than the theoretical value found to be $\SIE{4.24}{43}{\%}$. Finally the strongest $\gamma$-transitions in the decay of \Artt are shown including a line at $\SIE{4734}{3}{keV}$ associated to the decay of the IAS, which has not previously been identified.

\end{abstract}



\maketitle

\section{Introduction}
\label{introduction}
The decay of drip-line nuclei are distinguished \cite{KarstenRev} by having many open channels, due to the large $\beta$-decay energies and small particle-separation energies for these nuclei far from stability. This implies the decay can be used to study several different interesting topics. The many decay channels unfortunately also entail that the decay strength is more difficult to extract, since one will need a setup where all these channels can be detected. However, with a multi-hit detection setup, like the one described here, it is possible to study both the feeding from beta-decay and the structure of the proton-rich nuclei far from stability. 

\begin{figure*}[tbp]%
	\centering
	\includegraphics[width=1.90\columnwidth]{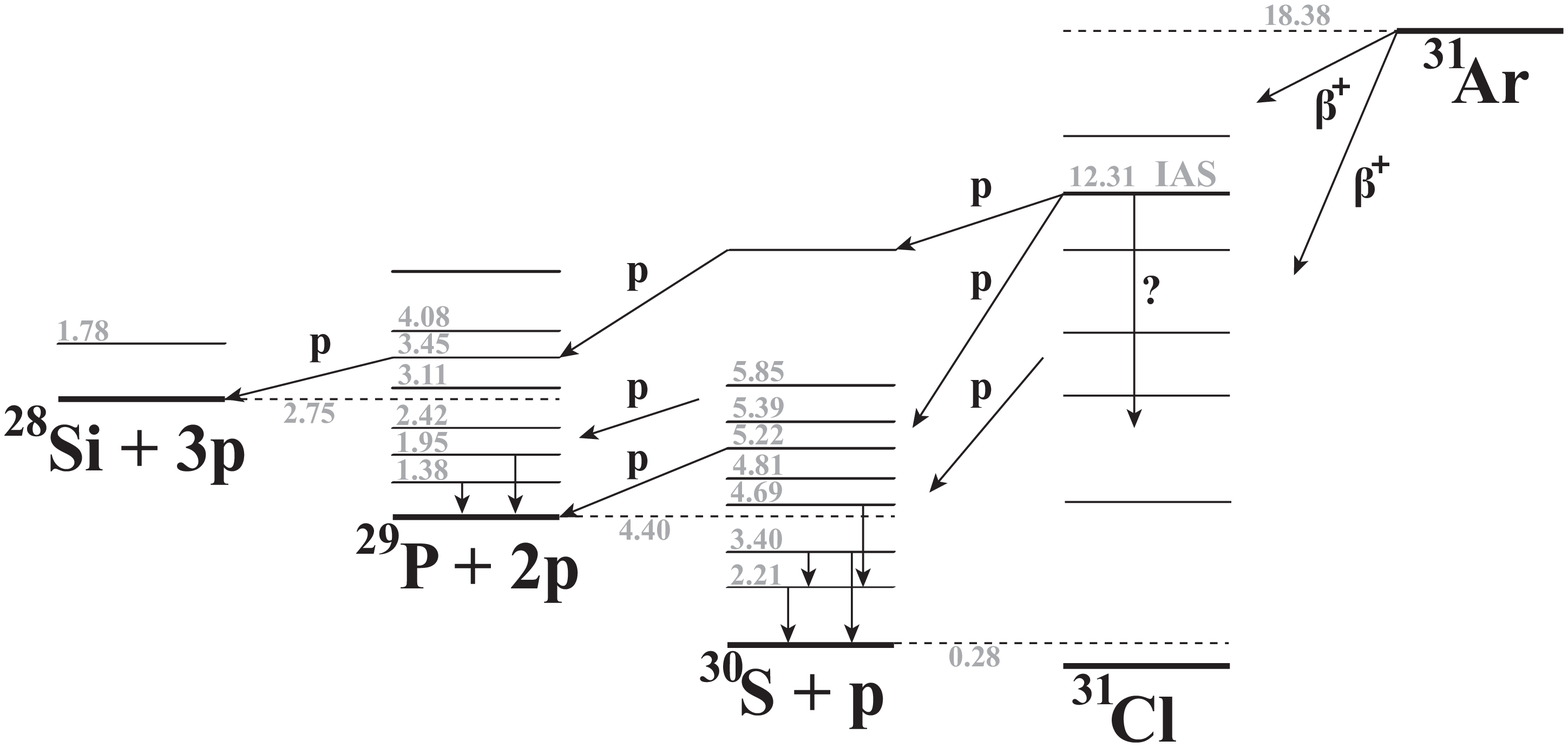} 
	\caption{The $\beta$ decay of \Ar, not to scale. Different proton and gamma decays are drawn as an illustration.}%
	\label{level}%
\end{figure*} 

The production is a challenge, since the further from stability the harder it is to produce nuclei with a sufficient yield. The proton-rich argon isotopes can be produced with a relative high yield and low contamination from CaO targets using the ISOL-technique. \Ar is therefore an ideal nucleus to use for this type of studies (the decay scheme of \Ar is shown in \fref{level}).  During the last few decades the decay of \Ar has been studied in several experiments at the ISOLDE radioactive ion beam facility at the European research organisation CERN. The first interest in this nucleus arose from the possibility of detecting a two-proton (2p) decay directly from the ground state of \Ar. Unfortunately this was not possible, but instead the $\beta$-delayed 2p-decay was measured \cite{Borge1990,Borrel87}. The mechanism of the $\beta$-delayed 2p-decay was studied in detail in two experiments at ISOLDE in 1995 \cite{Axelsson1998} and 1997 \cite{Fynbo2p} and found to be mainly sequential. A simultaneous component is predicted \cite{Brown90}, but there is still no experimental evidence for it. Our main current knowledge on the 
$\beta$-delayed 2p-emission stems from these two experiments studying the decay of \Ar. With the setup used in the experiment presented here, which had a high efficiency for proton detection with a good energy and angular resolution, the decay of \Ar can be used to study 
another exotic decay mode; the $\beta$-delayed 3p-emission, which has only been observed in two other nuclei so far; \ce{^{45}Fe} \cite{Fe3p} and \ce{^{43}Cr}  \cite{Cr3p,Audirac2012}. It was only recently discovered in the decay of \Ar by Pf\"utzner \emph{et al.} \cite{Ar3p}. The study of \Ar can thus now bring the same degree of information to this decay mode as it did to the $\beta$-delayed 2p-decay roughly 15 years ago.

A detailed knowledge on the $\beta$3p-decay will help assign the correct $\beta$-strength. Not only is it possible to measure the $\beta$3p-channels, but this will also make it possible to reassign decays that have previously been wrongly assigned as $\beta$2p-decays to the lowest states in \P, and thus assigned as $\beta$-decays to states in \Cl below the true ones. With a good detection efficiency for $\gamma$-rays, this can also be done by detecting all the particles of the delayed 2p$\gamma$-decay and in this way correctly identify the final state of the 2p-decay in \P.

Due to the sequential nature of the 2p-decay it can be used to study the levels in \S above the proton threshold, which are relevant for nuclear astrophysics. From the same experiment as discussed here, experimental limits on the ratio between the proton and gamma partial widths have been found for different low-lying levels in \S using the $\beta$2p-decay of \Ar \cite{no1}. This decay can also be used to determine the spin of the levels fed in the decay using proton-proton angular correlations. Until now only a tentative spin assignment has been made by comparison with the mirror nucleus \cite{Seto2013}.

A separate analysis of the Gamow-Teller strength using the $\beta$3p-decay of \Ar is in preparation \cite{no3}.

In the following section (\ref{theexperiment}) the experiment will be described. Section \ref{res-halflife} describes the new half-life determination of \Ar, then in section \ref{res-b3p} the spectroscopy analysis of the $\beta$3p-decay is presented followed by the analysis of the $\beta$2p$\gamma$-decay in section \ref{res-2pg}. The new improved results on the Fermi $\beta$-strength is given in section \ref{IASfeeding}. The new method for finding the spin of the low-lying levels in \S is presented in section \ref{spin} and used on the $\SI{5.2}{MeV}$ level, whose spin is still uncertain. Finally in section \ref{gamma33Ar} the $\gamma$-transitions in the decay of \Artt are presented, including a $\gamma$-line from the IAS, which is seen for the first time in the decay. In Section \ref{summary} the main results are summarised.



%
%
%
%
%
%

\section{The experiment}
\label{theexperiment}

\begin{figure}%
	\centering
	\includegraphics[width=0.7\columnwidth]{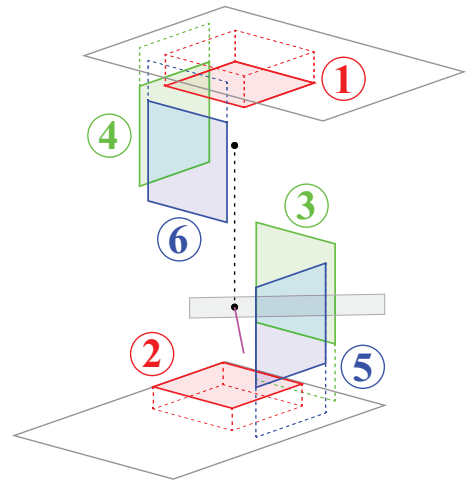}
	\caption{(Colour online) The experimental setup used for the experiment. The beam enters between DSSSD 5 and 6 and is stopped in a carbon foil mounted on a small metal holder entering between DSSSD 3 and 5. The top of the cube with three of the DSSSDs is lifted, following the dotted black line, for better visualisation. The two clustered germanium detectors that were situated outside the cube behind DSSSD 3 and 4, are not shown on the picture. }%
	\label{setupfig}%
\end{figure} 

The experiment was performed at the ISOLDE facility at CERN, Switzerland using the ISOL technique \cite{ISOLDE} with a powder CaO target and a versatile arc discharge plasma ion source \cite{CaOtarget}. The $\SI{60}{keV}$ ion beam was guided through the General Purpose Separator \cite{ISOLDE} to separate the desired argon isotope from background. However, a significant background from nitrogen (as \ce{N_{2}} and \ce{N_{2}H}) was present in the final beam. An average yield of \Ar of about 1 ion per second was obtained for a runtime of about 7 days. The beam was collected in a $\SI{50}{\mu g/cm^2}$ carbon foil situated in the middle of the Silicon Cube detector setup \cite{cube}. The Silicon Cube consists of six Double Sided Silicon Strip Detectors (DSSSDs) with backing in a cube formation, see \fref{setupfig}. For this experiment one detector with thickness of $\SI{69}{\mu m}$ (no. 1), one with a thickness of $\SI{494}{\mu m}$ detector (no. 5) and four detectors with a thickness close to $\SI{300}{\mu m}$ (no. 2, 3, 4, 6) were used, with $\SI{1500}{\mu m}$ thick $\SI{50}{mm}\times\SI{50}{mm}$ unsegmented silicon pad detectors used for backing to four of the detectors (no. 1, 2, 3, 6).

The geometry and energy calibration of the DSSSDs were made using a beam of \Artt produced from the same target-ion source unit as \Ar. A thorough description of the setup can be found in \cite{no1}. The energy calibration of the pad detectors behind the DSSSDs are made using a standard $\alpha$ calibration source (a \ce{^{148}Gd} source and a triple $\alpha$ source consisting of \ce{^{241}Am}, \ce{^{239}Pu}, and \ce{^{244}Cm}) without the DSSSDs present. The efficiency, $\epsilon_p$, is taken as the angular coverage, which for all the detectors is $\SIE{43}{2}{\%}$ of $4\pi$. 

Two Miniball germanium cluster detectors \cite{Miniball} were situated outside the cube chamber behind detector 3 and 4. Each detector consists of three clusters, but unfortunately one of the clusters of the detector behind DSSSD 3 gave no signal. A preliminary energy calibration of the clusters was made using \ce{^{137}Cs} and \ce{^{60}Co}. This was then improved using a \ce{^{152}Eu} source together with high energy $\gamma$-lines from the decay of \ce{^{16}N} and \ce{^{15}C}, which were found in the runs with \Ar. This gives an energy calibration up to $\SI{1.8}{MeV}$ with an uncertainty of $\SI{1}{keV}$. Above this energy it was found, using decays of \ce{^{16,18}N} and \ce{^{32,33}Ar} recorded online, that the energy should be shifted upward by $\SI{0.7}{keV}$. Doing this gives an uncertainty of $\SI{1}{keV}$ for energies between $\SI{1.8}{MeV}$ and $\SI{2.5}{MeV}$. For energies above this the uncertainty is $\SI{3}{keV}$.

A total efficiency calibration was made for the two Miniball detectors. First an absolute efficiency calibration was made using the low-lying $\gamma$-lines from a \ce{^{133}Ba} source with a known activity of $\SIE{17.0}{3}{kBq}$ at the time of the experiment. The detection efficiency for the $\gamma$-lines from the \ce{^{152}Eu} source, corrected for emission probabilities using \cite{Debertin}, is then scaled, using the $\SI{302}{keV}$- and $\SI{356}{keV}$-points from \ce{^{133}Ba} and placing the $\SI{344}{keV}$-point from \ce{^{152}Eu} on a straight line between these. The absolute $\gamma$ efficiency above $\SI{600}{keV}$ was then found by fitting the \ce{^{152}Eu} points to a relative efficiency curve determined in a slightly different detector configuration \cite{Jacob2013} (that used four different $\gamma$ sources: \ce{^{152}Eu}, \ce{^{60}Co}, \ce{^{207}Bi} and \ce{^{11}Be}). The result, using the formula in Ref. \cite{miniballeff}, is
\begin{align}
	\varepsilon_{\gamma} \left( E\right)  =& 0.21 \exp\Bigg(-2.669 - 1.457\log\left( \frac{E}{\SI{}{MeV}}\right) \nonumber\\ 
	 &- 0.231 \left[\log\left( \frac{E}{\SI{}{MeV}}\right) \right]^2  \Bigg) ,
\end{align}
with an estimated uncertainty of $\SI{10}{\%}$.

For normalisation of the total number of \Ar collected during the run, the largest one-proton peak at $\SI{2083}{keV}$ with an absolute branching ratio of $\SIE{26.2}{29}{\%}$ \cite{Axelsson98b} is used. A small fraction of the activity could only be seen from the beam entrance side. Furthermore, the target holder shadows several pixels in particular for detector 1 and 2. These effects are all included in the detailed Monte Carlo simulations used below to extract final branching ratios. An overall estimate of the number of \Ar collected during the experiment is $\num{5.6(6)e5}$.

\section{Results and discussion}
\label{results}

\subsection{Half-life of \Ar}
\label{res-halflife}
The half-life of \Ar is found in the same way as in \cite{Fynbo2p}. We used only the data recorded after the beam gate was closed at $\SI{100}{ms}$ after proton impact on the production target. Only the strongest 1p-peak at $\SI{2083}{keV}$, corresponding to an energy range between $\SI{2040}{keV}$ and $\SI{2120}{keV}$, was used to eliminate background. In this way the data could be fitted using the maximum likelihood method to a single exponential component and a constant background. This gave a half-life of $\SIE{15.1}{3}{ms}$, which is consistent with previous determinations of $\SIE{14.1}{7}{ms}$ \cite{Fynbo2p}, $\SIE{15}{3}{ms}$ \cite{Borrel87} and $15.1^{+1.3}_{-1.1}\SIU{ms}$ \cite{Bazin92}.

\subsection{$\beta$-delayed three-proton spectroscopy}
\label{res-b3p}
\label{beta3p}
\begin{figure}%
	\centering
	\includegraphics[width=0.90\columnwidth]{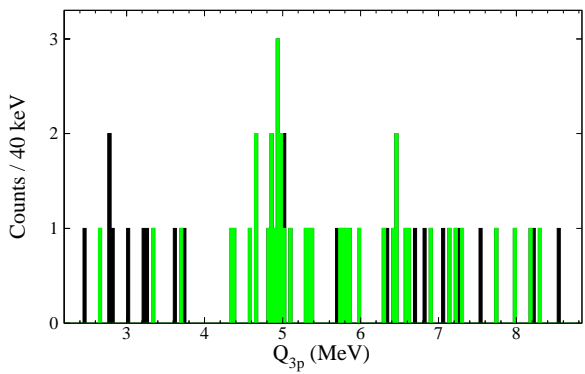} 
	\caption{(Colour online) \Qep for three-particle events. The histogram shows all the events where the first two particles detected have energies above $\SI{800}{keV}$ and the third has an energy above $\SI{500}{keV}$ unless it is in detector 5, where it has to have an energy above $\SI{800}{keV}$. The part of the histogram that is green/grey is without the events that are most likely not real 3p-events, which are removed as described in the text.}%
	\label{q3pfig}%
\end{figure} 

Here we present the first spectroscopic analysis of a $\beta$-delayed 3-proton decay. A spectrum of the \Qep-values calculated for the 3p-events observed during the experiment is shown in \fref{q3pfig}. To eliminate noise and contamination from $\beta$-particles the following energy gates are used: The energy of the first two particles detected should exceed $\SI{800}{keV}$ and the energy of the third $\SI{500}{keV}$ unless it is in the thick detector (detector 5), where a $\beta$-particle will deposit more energy. In this case the gate is set at $\SI{800}{keV}$. The reason for allowing the third particle to have an energy less than the others is that the 3p-decay could go through a $\frac{7}{2}^-$ level in \P $\SI{699}{keV}$ above the proton threshold. In principle the decay can also go trough a $\frac{5}{2}^+$ level only $\SI{357}{keV}$ above the proton threshold. However, the penetrability for this level is roughly a factor of $25$ below the penetrability for the $\frac{7}{2}^-$ level. Furthermore, it is not possible with our setup to distinguish these low-energy protons from $\beta$-particles. We thus first assume that there are no transitions through the $\frac{5}{2}^+$ level. In the end of this section we will return to this issue and argue that this is a good assumption.


From the $\beta$-delayed proton decay of \Artt it is found that the probability of detecting the $\beta$-particle when a proton is detected, using the same energy cuts as used for the first and last particle above, is $\SIE{0.43}{5}{\%}$. Using this probability for detecting a $\beta$-particle having a real 2p-event in \Ar, one finds that 29(4) of the 62 detected three-particle events are presumably 2p$\beta$-events. Further cuts are therefore introduced to reduce this background: $\beta$-particles often scatter in the detector, so events with three or more hits in the same detector with approximately the same energy, can be discarded. There are also two events with \Qep$>\SI{13}{MeV}$, which cannot be a real 3p-event, since there is only $\SIE{10.9}{2}{MeV}$ available for the $\beta$3p-decay of \Ar \cite{mass}. 
Furthermore, if the two particles with the highest energy have a \Qtp-value corresponding to the known decays of the Isobaric Analogue State (IAS) to one of the three lowest states in \P (see Section \ref{IASfeeding} or \fref{fynbofig}) and the energy of the third particle is less than $\SI{1.2}{MeV}$, then these are most likely 2p$\beta$-events. These events should thus also be discarded. In this way 21 events are removed and the remaining can be seen as the green/grey histogram in \fref{q3pfig}. In the following only the 41 remaining events are considered. 

A broad peak is seen in the green/grey histogram of \fref{q3pfig} around $\SIE{4.89}{29}{MeV}$ containing 19 events (between $\SI{4.3}{MeV}$ and $\SI{5.5}{MeV}$). To investigate the spread in \Qep due to detection resolution a simulation was made, which showed that the expected full width half maximum is more than $\SI{300}{keV}$. A real peak of events from a given level in \Cl is thus expected to be as broad as the one at $\SIE{4.89}{29}{MeV}$. This peak is most likely due to the 3p-decay of the IAS, since it corresponds to a \Cl-level at an energy of $\SIE{12.32}{29}{MeV}$. It is interesting to notice that it is only approximately half of the 3p-events that belong to the decay of the IAS. The other half stem from transitions from levels in \Cl above the IAS. Due to the large $Q$-window for particle emission these levels will most likely not decay to the ground state in \P and it can thus be difficult to make a correct assignment of these if only two of the protons are detected, as is the case for previous studies of the \Ar decay. Using now the $\beta$3p-decay a better determination of the Gamow-Teller strength can be found. This will be published separately in \cite{no3}. 

\begin{figure}%
	\centering
	\includegraphics[width=0.90\columnwidth]{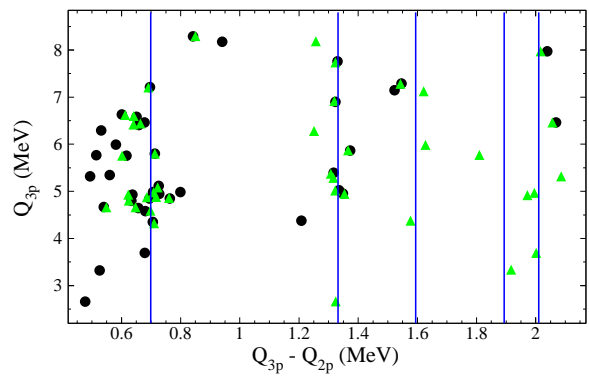} 
	\caption{(Colour online) \Qep vs. $Q_{3p} - Q_{2p}$. The lines indicates the levels in \P. For the black circles \Qtp is calculated assuming that the first two particles are the ones with the highest energy and for the green/grey triangles it is calculated to best fit the five levels in \P shown.}%
	\label{3pdifffig}%
\end{figure} 

\begin{figure}%
	\centering
	\includegraphics[width=0.90\columnwidth]{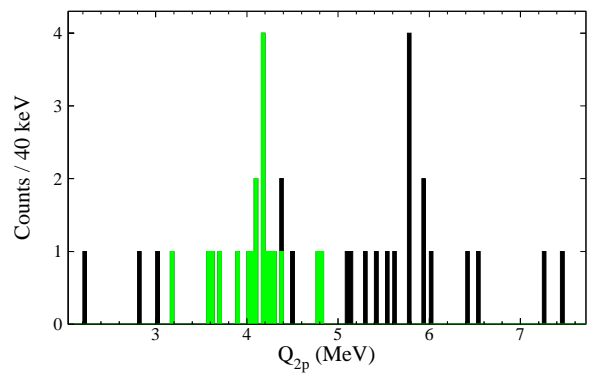} 
	\caption{(Colour online) \Qtp calculated from the two particles with highest energy. The histogram is from the 41 events in the green/grey histogram of \fref{q3pfig}. The part that is green/grey is the 19 events around $\SIE{4.84}{29}{MeV}$ in the green/grey histogram of \fref{q3pfig}.}%
	\label{3pq2pfig}%
\end{figure} 

The decay mechanism of the $\beta$-delayed 3p-decay is unknown. If it is sequential it should be possible to identify it going through different levels in both \S and \P. However, due to limited statistics and the density of levels for high excitation energies in \S it will not be possible to do this for \S. If the decay goes through levels in \P these can be identified via the difference between the \Qep-value and the \Qtp-value, since this difference then corresponds to the energy of the level populated in \P minus the proton separation energy. The reason to use the two $Q$-values is that these can be extracted directly from the experimental data and that a correction for the recoil of the daughter nucleus is included. The \Qep-value can be calculated independently on the decay mechanism, while for the \Qtp-value one must chose which particles should be considered to be the first two in the decay. In \fref{3pdifffig} this difference is plotted for two different choices together with lines indicating the levels in \P. For the black dots it is assumed that the first two particles are the ones with the highest energy. This is, however, not necessarily a reasonable assumption for all the events. Instead the first two particles can be chosen so that the difference between the \Qep- and \Qtp-value fits the levels in \P (only the first five levels were chosen). This choice is plotted as the green/grey triangles. However, due to the large expected spread in the calculated \Qep-value, a wrong assignment can easily be made. So even though the 3p-decay seen here is fully consistent with being sequential, a large contribution from direct decay cannot be excluded. When the energy of all three particles are above $\SI{1.2}{MeV}$, they are most likely all protons, but the density of states in \P is so high here compared to the expected spread that it is easy to interpret a direct decay as a sequential decay. This is not a problem for the level at $\SIE{3447.6}{4}{keV}$ (corresponding to a difference between \Qep and \Qtp of $\SI{699}{keV}$). The problem here is that one of the particles has an energy around $\SI{0.7}{MeV}$ and it is thus difficult to distinguish protons from $\beta$-particles. The majority of these events stem from the peak in the \Qep-spectrum around $\SIE{4.89}{29}{MeV}$, see \fref{q3pfig}, that most likely belongs to the decay of the IAS. Their \Qtp-value can be seen in \fref{3pq2pfig}: More than half of them lie around $\SIE{4.14}{13}{MeV}$. Assuming they go through the $\SIE{3447.6}{4}{keV}$-level in \P, this corresponds to a \Cl-energy at $\SIE{12.27}{13}{MeV}$ in complete agreement with the value of $\SIE{12.32}{29}{MeV}$ from the \Qep-value. These events cannot be 2p$\beta$-events. If they were, one would expect more than $\num{2e3}$ events at this energy in the \Qtp-spectrum made from 2p-events. There are indications of small peaks around this energy, but they contain less than 70 events. The IAS decay can thus be assumed to go through the level at $\SIE{3447.6}{4}{keV}$ in \P, which supports the theory that the events lying close to $\SI{699}{keV}$ in \fref{3pdifffig} are also events going through this level.

We also have strong indications of events going through the level at $\SIE{4080.5}{3}{keV}$ (corresponding to a difference between \Qep and \Qtp of $\SI{1332}{keV}$). With the statistics available here and the expected large spread in the \Qep, this is, however, not conclusive. The sparsity of events with $Q_{3p} - Q_{2p}$ between $\SI{0.9}{MeV}$ and $\SI{1.1}{MeV}$ is a strong indication that there are no simultaneous 3p-decays with a low-energy proton.

We now return to the issue of possible involvement of the $\frac{5}{2}^+$ level at $\SIE{3105.9}{3}{keV}$ level $\SI{357}{keV}$ above the proton threshold. By using measurements of the ressonance strength, $(2J+1)\frac{\Gamma_p\Gamma_{\gamma}}{\Gamma}$, and the lifetime \cite{NDS29} one finds for the $\frac{7}{2}^-$ level that $\Gamma\sim\Gamma_{\text{max}}=\SIE{51}{31}{meV}$ and $\Gamma_{\text{min}}=\SIE{0.038}{10}{meV}$, where $\Gamma_{\text{max}}$ ($\Gamma_{\text{min}}$) refers to the largest (smallest) width of $\Gamma_p$ and $\Gamma_{\gamma}$. For the $\frac{5}{2}^+$ level one finds $\Gamma\sim\Gamma_{\text{max}}=\SIE{19}{9}{meV}$ and $\Gamma_{\text{min}}=\SIE{0.46}{11}{meV}$. If $\Gamma_{\text{max}} = \Gamma_{\gamma}$ for the  $\frac{7}{2}^-$ level, one would expect to see around $700$ $\gamma$-rays at $\SI{1493.6}{keV}$, corresponding to the decay of this level to the second excited level, when gating on two protons. This we do not see in our two-proton gated $\gamma$-spectrum, see \sref{res-2pg} and \fref{gam2pfig}. We therefore conclude that $\Gamma_p=\Gamma_{\text{max}}$. Looking now at the mirror nucleus \ce{^{29}Si}, where the $\frac{7}{2}^-$ and $\frac{5}{2}^+$ level both lies below the proton threshold, we see that the half-lives of these two levels are $\SIE{2.63}{9}{ps}$ and $\SIE{33}{1}{fs}$ respectively. The half-lives of the two levels in \P are $\SIE{9}{6}{fs}$ and $\SIE{23}{10}{fs}$ respectively. By comparison it is reasonable to assume that $\Gamma_p=\Gamma_{\text{max}}$ for the $\frac{7}{2}^-$ level, as deduced above, and $\Gamma_p=\Gamma_{\text{min}}$ for the $\frac{5}{2}^+$ level. From this it is found that the proton width of the $\frac{5}{2}^+$ level is $111(72)$ times smaller than the proton width of the $\frac{7}{2}^-$ level and it is thus reasonable to assume that the 3p-decay through the $\frac{5}{2}^+$ level is suppressed.


\subsection{$\beta$-delayed 2p$\gamma$-decay}
\label{res-2pg}

\begin{figure}%
	\centering
	\includegraphics[width=0.90\columnwidth]{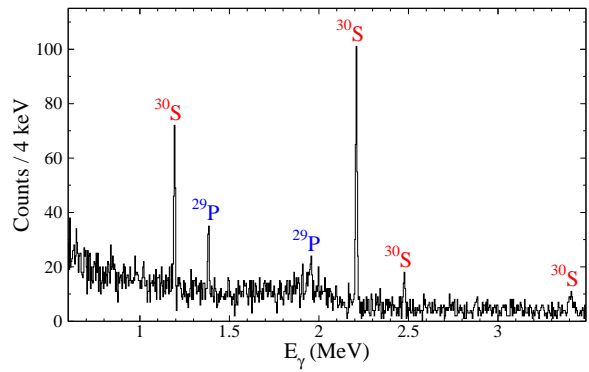} 
	\caption{The $\gamma$-spectrum gated on one proton with an energy above $\SI{800}{keV}$.}%
	\label{gam1pfig}%
\end{figure} 

\begin{figure}%
	\centering
	\includegraphics[width=0.90\columnwidth]{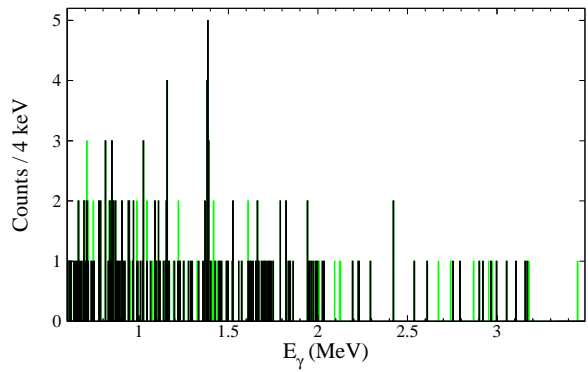} 
	\caption{(Colour online). The $\gamma$-spectrum gated on two protons. For the black spectrum both protons have energies above $\SI{800}{keV}$. The extra events in green/grey are $gamma$-rays where one of the protons has an energy between $\SI{500}{keV}$ and $\SI{800}{keV}$ and has not hit detector 5.}%
	\label{gam2pfig}%
\end{figure} 

The indications of a sequential 3p-branch implies that the decay populate higher-lying levels in \P than previously found. With our experiment it was possible to detect $\gamma$-rays in coincidence with protons and we thus have a chance to see the $\gamma$-transitions from these levels for the first time. However, the detection efficiency is limited and the chance of detecting the $\gamma$-ray in coincidence with both the emitted protons is thus very small. For a real 2p$\gamma$-event it is $2/\epsilon_p=4.6$ 
times more likely to detect it as a 1p$\gamma$-event than to detect it as a 2p$\gamma$-event. So to search for the transitions from higher-lying levels in \P one should start by considering the one-proton gated $\gamma$-spectrum, which is shown in \fref{gam1pfig}. As previously $\SI{800}{keV}$ is used as a lower energy cut on the proton. In this spectrum clear peaks are identified from the lowest states of both \S and \P, see \cite{no1}. But due to background in the spectrum there are no clear signal from the levels above the second excited state in \P. In \fref{gam2pfig} the two-proton gated $\gamma$-spectrum is shown. Two different gates are used: One where both particles have energies above $\SI{800}{keV}$ (black) and one where the second particle has an energy above $\SI{500}{keV}$ unless it is in the thick detector 5, then it has an energy above $\SI{800}{keV}$ (black + green/grey). In the following all the levels in \P up to $\SI{4.1}{MeV}$ will be considered and the number of 2p$\gamma$-events will be compared with the one expected from the 1p-gated $\gamma$-spectrum. Since there is no reason why the second emitted proton should have an energy above $\SI{800}{keV}$ instead of just $\SI{500}{keV}$, the extra events in green/grey in \fref{gam2pfig} is also included.

The first excited state in \P ($\frac{3}{2}^+$) is at $\SIE{1383.55}{7}{keV}$ \cite{NDS29}. It decays to the ground state and the peak is clearly seen in both the 1p- and 2p-gated $\gamma$-spectra. There are $64(11)$ events above background in the 1p-gated spectrum. This implies that there should be $14(2)$ events in the 2p-gated spectrum, which agrees very well with the $13(4)$ measured above background.

The second excited state at $\SIE{1953.91}{17}{keV}$ ($\frac{5}{2}^+$) decays primarily to the ground state. A peak at this energy is seen in the 1p-gated $\gamma$-spectrum. It contains $59(15)$ events, but it is more than twice as broad as the other peaks in the spectrum. This and the discussion in \cite{no1} indicates that there might be other contributions to the peak. From the $59(15)$ events in the 1p-gated spectrum one would expect $13(3)$ events in the 2p-gated spectrum. Only $7(3)$ events are observed in total, but if there are other contributions to the peak in the one-proton gated spectrum the expected number would be smaller.

The third excited state is a $\frac{3}{2}^+$ state at $\SIE{2422.7}{3}{keV}$, and decays also primarily to the ground state. There is no significant signal above background in the 1p-gated $\gamma$-spectrum at this energy. In the 2p-gated spectrum there are two events with no significant background at $\SIE{2422}{11}{keV}$. This would imply $9(7)$ events in the 1p-gated spectrum. Considering the background level in this area in the 1p-gated spectrum it is not possible to disprove this.

The next level is the $\SIE{3105.9}{3}{keV}$-level, which is just above the proton threshold. It is a $\frac{5}{2}^+$-level and decays primarily by a $\SI{1722.2}{keV}$ $\gamma$-ray. Again there is no significant signal above background in the 1p-gated spectrum. There are a maximum of $14(9)$ events above background, which implies there should be $3(2)$ events in the 2p-gated spectrum, where there are a total of $2$. 

The $\frac{7}{2}^-$-level at $\SIE{3447.6}{4}{keV}$, which was identified in the 3p-decay, has a total half-life of $\SIE{9}{6}{fs}$. It decays primarily by a $\SI{1493.6}{keV}$ $\gamma$-ray. There is a hint of a peak in the 1p-gated $\gamma$-spectrum at this energy containing $14(7)$ events. From this one expects $3.0(15)$ events in the 2p-gated spectrum, where there are a total of $2$.

The level at $\SIE{4080.5}{3}{keV}$ is a $\frac{7}{2}^+$ level with a total half-life of $\SIE{11}{1}{fs}$. It decays primarily by a $\SI{2126.3}{keV}$ $\gamma$-ray. In the 1p-gated $\gamma$-spectrum, there is no indications of a peak at this energy. There are $5(5)$ events, which means that there should be $1(1)$ event in the 2p-gated spectrum and there are a total of $2$ events.

So in summary, the $\gamma$-decays observed for the six lowest levels in \P give consistent results for the 1p- and 2p-gated $\gamma$-spectra, but only the feeding of the lowest two can be seen directly in the $\gamma$-spectra.

\subsection{The Fermi strength of the $\beta$-decay}
\label{IASfeeding}
The Fermi strength of the $\beta$-decay of \Ar has been measured previously \cite{Fynbo2p}, by considering the 1p- and 2p-decay of the IAS to the lowest states in \S and \P. As shown in \cite{no1} and in the results presented above for the IAS, this is not sufficient. In addition to the 3p-decay, all the levels up to the proton threshold should in principle be included for the 1p- and 2p-decays and also the levels just above the proton threshold, since they are not considered, due to the lower energy cut on the proton, in the 2p- and 3p-decays. To get a precise determination of the branching ratios for the different decays it is important to use spectra with a good energy resolution and to precisely know the total number of \Ar collected and the efficiency for detection. For this reason only the $\SI{300}{\mu m}$ DSSSD's with backing can be used. Detector 2 had several broken strips and less accurate efficiency determination due to shading from the target holder. This leaves only detector 3 and 6, which are used to determine the branching ratios for the two- and one-proton decays. The statistics is so low for the three-proton decay that all the detectors are needed, and the branching ratio is thus found using the data presented in Section \ref{beta3p}. It is listed in \tref{IAStab} together with the branching ratios found for the two- and one-proton decays. The branching ratio found here for the three-proton decay to the ground state of \Si is consistent with the $\SI{99}{\%}$ confidence upper limit of $\SI{0.11}{\%}$ found by Fynbo \emph{et al.} \cite{Fynbo3p}. 


\begin{figure}%
	\centering
	\includegraphics[width=0.90\columnwidth]{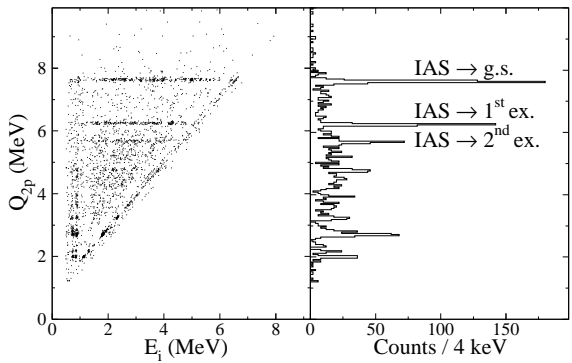} 
	\caption{Two-proton spectrum made using only detector 3 and 6 with a lower cut-off of $E>\SI{500}{keV}$. Left: \Qtp vs. the energy, $E_i$, of the two particles. Right: The projection onto the \Qtp-axis.}%
	\label{q2p36fig}%
\end{figure} 

The two-proton spectrum using only detector 3 and 6 (with $E>\SI{500}{keV}$) can be seen in \fref{q2p36fig}. The peaks corresponding to the decay to the ground state and the first and the second excited state of \P are clearly identified in the spectrum at \Qtp-values at $\SI{7.6}{MeV}$, $\SI{6.3}{MeV}$ and $\SI{5.7}{MeV}$. There is a small indication of a peak at $\SI{5.2}{MeV}$ corresponding to the transition to the third excited state in \P. Transitions to higher-lying states can not be identified. The values given here (in \tref{IAStab}) are all lower than reported by  Fynbo \emph{et al.} \cite{Fynbo2p}. The main reason for this is that the energy and angular resolution in this experiment is better for this energy range and thus our peaks are significantly narrower. 
This means that Ref. \cite{Fynbo2p} includes contributions from decays with \Qtp-values close to those for the IAS decays. Furthermore, for the decays to excited states in \P, the background from Gamow-Teller transitions is estimated and subtracted here, which is not done in Ref. \cite{Fynbo2p}.

\begin{figure}%
	\centering
	\includegraphics[width=0.90\columnwidth]{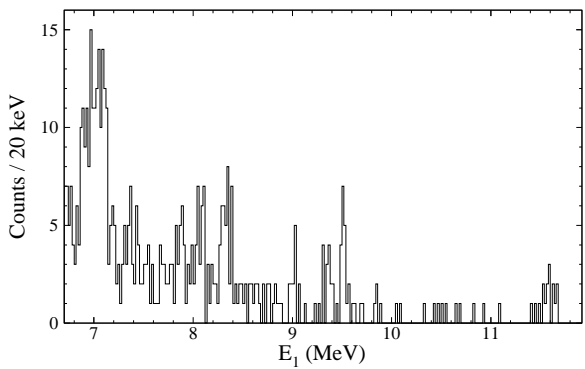} 
	\caption{The one-proton energy spectrum for detector 3 for high energies.}%
	\label{1p3fig}%
\end{figure} 

The one-proton energy spectrum for detector 3 can be seen in \fref{1p3fig} (the spectrum for detector 6 is similar). The branching ratios are found separately for detector 3 and 6 and the average is given in \tref{IAStab}. The large uncertainty in the energy is due to limited statistics and a large uncertainty in the calibration of the back detectors for high energies. Since the energy cut-off on the two-proton spectra is $\SI{500}{keV}$ the branching ratios are found up to the \S-level at $\SIE{4809.0}{3}{keV}$ \cite{no1} ($\SI{413}{keV}$ above the proton threshold) in the one-proton spectrum. The peaks at $\SI{8.1}{MeV}$ and $\SI{7.0}{MeV}$ cannot be separated into two components, even though they should both contain contributions from decays to two different levels in \S. The branching ratios are thus found for the total contribution from the two levels.

If we neglect isospin symmetry breaking and disregard Gamow-Teller contributions the $\beta$-strength to the IAS is $B_F=5$. Using the Coulomb displacement energy of \ce{^{32,33,34,35}Ar} extracted from Ref. \cite{ADNDT99} we estimate the Coulomb displacement energy of \Ar to be $\SIE{6.85}{10}{MeV}$ giving $Q_{\text{EC}}=\SIE{18.38}{10}{MeV}$. With this and our improved half-life of \Ar (see Section \ref{res-halflife}) we obtain a total theoretical branching ratio of $\SIE{4.24}{43}{\%}$, where the large uncertainty stem from the uncertainty on the $Q_{\text{EC}}$-value. Without this the uncertainty of the total theoretical branching ratio is only $\SI{0.09}{\%}$. A better determination of the mass of \Ar would thus be very beneficial. The theoretical branching ratio is larger than the experimental value of $\SIE{3.60}{44}{\%}$, but within one standard deviation. The uncertainty on the experimental value cited does not include the relatively large uncertainty stemming from the normalisation of the number of \Ar ions in the experiment using the absolute branching ratio of the main 1p-peak (see Section \ref{theexperiment}). The results here are an improvement to the earlier result by Fynbo \emph{et al.} \cite{Fynbo2p}, but note that the uncertainties on both the total experimental and theoretical branching ratios quoted there are underestimated. However, there remain levels in \P below the proton threshold and one above to which two-proton decays could not be extracted. We could also not identify any $\gamma$-rays corresponding to transitions in \Cl from the IAS, but a contribution from these cannot be excluded, since $\gamma$-transitions from the IAS have been found in the decay of both \ce{^{32}Ar} \cite{Ar32gamma} and \Artt (see Section \ref{gamma33Ar}).

\begin{table*}[tbp]
\caption{Branching ratios for the decay of the IAS. The \Cl-energies are found using the masses from \cite{mass} and a proton separation energy for \Cl of $\SIE{282.8}{44}{keV}$ \cite{AnttiPhd}. The decays written in \textit{italic} correspond to decays not uniquely identified in the spectra: There is marginal indication of the two-proton branch and the one-proton branches cannot be uniquely assigned to levels in \S. The total branching ratio is quoted with and without these decays. The efficiencies used are different for each of the three decay modes and the uncertainty stemming from these are included in the cited uncertainties for each decay. The correlation is taken into account for the uncertainty on the total branching ratio. Furthermore, there is a systematic error of $\SI{11}{\%}$ stemming from the normalisation (See Section \ref{theexperiment}), which is not included in the cited uncertainties.}
\begin{tabular}{c c c c c}
\toprule[0.1em]
\vspace{-1mm} \\
\multicolumn{5}{c}{Three-proton branch}
\vspace{2mm} \\
\midrule[0.08em]
 \vspace{-2mm} \\
 \vspace{0.8mm}
Final state in \Si (keV)	& $J^{\pi}$	& \Qep (MeV) 	& $E_{\text{IAS}}$ (MeV)		& B.R. (\%) \\
\midrule[0.08em]
\vspace{-3mm} \\
\vspace{0.4mm}0		&	$0^+$		&	$4.89(29)$	& $12.32(29)$							& $0.039(19)$ \\ 
\midrule[0.1em]
\vspace{-1mm} \\
\multicolumn{5}{c}{Two-proton branch}
\vspace{2mm} \\
\midrule[0.08em]
 \vspace{-2mm} \\
 \vspace{0.8mm}
Final state in \P (keV)	& $J^{\pi}$				& \Qtp (MeV) 	&$E_{\text{IAS}}$ (MeV)		& B.R. (\%) \\
\midrule[0.08em]
\vspace{-3mm} \\
0									&	$\frac{1}{2}^+$		&$7.633(4)$		& $12.311(6)$					& $1.47(23)$ \\ 
1383.55(7)					&	$\frac{3}{2}^+$		&$6.251(4)$		& $12.313(6)$					& $0.88(15)$ \\ 
1953.91(17)					&	$\frac{5}{2}^+$		&$5.688(6)$		& $12.320(8)$					& $0.40(10)$ \\ 
\vspace{0.4mm}\textit{2422.7(3)}	&	$\mathit{\frac{3}{2}^+}$	&\textit{5.22(8)}		& \textit{12.32(8)}					& \textit{0.075(50)} \\
\midrule[0.1em]
\vspace{-1mm} \\
\multicolumn{5}{c}{One-proton branch}
\vspace{2mm} \\
\midrule[0.08em]
 \vspace{-2mm} \\
 \vspace{0.8mm}
Final state in \S (keV)	& $J^{\pi}$	& $E_p$ (MeV) 	&$E_{\text{IAS}}$ (MeV)		& B.R. (\%) \\
\midrule[0.08em]
\vspace{-3mm} \\
0									&	$0^+$		&	$11.57(8)$	& $12.24(8)$					& $0.049(11)$ \\ 
2210.2(1)						&	$2^+$		&	$  9.46(8)$	& $12.27(8)$					& $0.104(18)$ \\ 
3404.1(1)						&	$2^+$		&	$  8.33(8)$	& $12.30(8)$					& $0.108(17)$ \\ 
\textit{3667.7(3)}						&	$\mathit{0^+}$		& \multirow{2}{*}{\textit{ 8.08(8)}} & \textit{12.30(8)}	& \multirow{2}{*}{\textit{0.101(21)}}  \\
\textit{3677.0(3)}					&	$\mathit{1^+}$		&						& \textit{12.31(8)}				&						\\
\textit{4687.7(2)}						&	$\mathit{3^+}$		& \multirow{2}{*}{\textit{ 7.01(8)}} & \textit{12.22(8)}	& \multirow{2}{*}{\textit{0.38(4)}}  \\
\vspace{0.4mm}\textit{4809.0(3)}					&	$\mathit{2^+}$		&						& \textit{12.34(8)}				&						\\
\midrule[0.1em]
 \vspace{-2mm} \\
\vspace{0.8mm}Total			&					&						& $12.313(4)$					& $3.05(42)$ \\ 
\vspace{0.8mm}\textit{Total}			&					&						& 					&\textit{3.60(44)}\\
\bottomrule[0.1em]
\end{tabular}
\label{IAStab}
\end{table*}



\subsection{Spin of low-lying levels of \S}
\label{spin}

A detailed knowledge of the levels just above the proton threshold in \S is important for determining the reaction rate of \rec, which influences the silicon abundances, which can be directly studied from presolar dust grains believed to be produced in classical novae. In the last few years the relevant levels in \S have been studied intensively \cite{no1,Seto2013,Lotay2012,Almaraz2012,Richter2013}, such that the energies are now known for the relevant levels, while some disagreements about the spin assignment remain. In this section we will present a new method for determining the spin of these levels. The method will be used to give the first determination of the spin of the $\SI{5.2}{MeV}$-level populated in the \Ar decay.  

The spin of the low-lying levels of \S can be found using
proton-proton angular correlations in 2p-decays through the
interesting levels.  The distribution of angles, $\theta$, between the
two protons can be written as \cite{Bie53}
\begin{align*}
	W(\cos\theta) = \sum_{\nu=0}^{\nu_{\text{max}}} A_{\nu}P_{\nu}\left(\cos\theta\right),
\end{align*}
where $P_{\nu}$ is the $\nu$'th Legendre Polynomial and the sum
extends to
\begin{align*}
	\nu_{\text{max}} = \text{min}\left(2l_1,2l_2,2j\right),
\end{align*}
so that one obtains an isotropic distribution if the angular momenta
involved are too small.
Here $j_1$ ($l_1$) and $j_2$ ($l_1$) are the spin (orbital angular momentum) of the first and second emitted
proton, respectively, and $j$ is the spin of the \S state coupled with
the first proton. The coefficient $A_{\nu}$ is given by
\begin{align*}
	A_{\nu} &= F_{\nu}\left(l_1,j_1,j\right)b_{\nu}\left(l_1,l_1\right)F_{\nu}\left(l_2,j_2,j\right)b_{\nu}\left(l_2,l_2\right) \\
	b_{\nu}\left(l,l'\right) &= \frac{2\sqrt{l(l+1)l'(l'+1)}}{l(l+1)+l'(l'+1)-\nu(\nu+1)},
\end{align*}
where $F_{\nu}$ can be found from the tabulation in \cite{Bie53}.

In the $\beta$2p-decay of \Ar we expect the excess protons to be
mainly in the $sd$-shell and shall therefore make the assumption that
only positive parity states in \S will be populated. The possible
values for $A_{2}$ in the decay are given in \tref{A2tab}. In many
cases there are two possible values for $j$ and the table indicates
the range spanned by the two extreme situations in which only one $j$
value contributes.

\begin{table}[bthp]
\caption{The $A_2$ coefficients for 2p-transitions calculated for the
  different initial states, $J^{\pi}_i$, (in \Cl) through five positive
  parity states, $J^{\pi}_m$, (in \S) to a $\frac{1}{2}^+$ final state
  (ground state of \P).}
\begin{tabular}{c | c c c}
\toprule[0.1em]
\vspace{-3mm} \\
\diagbox{$J^{\pi}_m$}{$J^{\pi}_i$} & $\frac{3}{2}^+$ &   $\frac{5}{2}^+$  & $\frac{7}{2}^+$     \\
\midrule[0.05em]
\vspace{-3mm} \\
$0^+$ & 0 & 0 & 0 \\
$1^+$ & 0 & 0 & 0 \\
$2^+$ & 0 & 0 & $[-0.70;-0.25]$ \\
$3^+$ & $[0.15;0.87]$  & 0 & 0 \\
\vspace{0.4mm}$4^+$ & $[0.76;1.00]$  & $[0.13;0.95]$  & 0 \\
\bottomrule[0.1em]
\end{tabular}
\label{A2tab}
\end{table}

\begin{figure}%
	\centering
	\includegraphics[width=0.90\columnwidth]{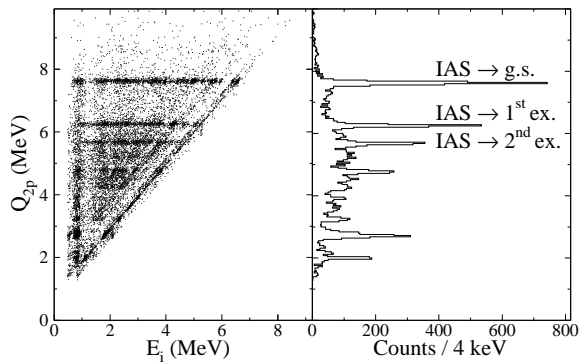}
	\caption{Two-proton spectrum with lower cut-off: $E_1>\SI{800}{keV}$ and $E_2>\SI{500}{keV}$, except for detector 5, where $E_2>\SI{800}{keV}$. Left: \Qtp vs. the energy, $E_i$, of the two particles. Right: The projection onto the \Qtp-axis.}%
	\label{fynbofig}%
\end{figure} 

\begin{figure}%
	\centering
	\includegraphics[width=0.90\columnwidth]{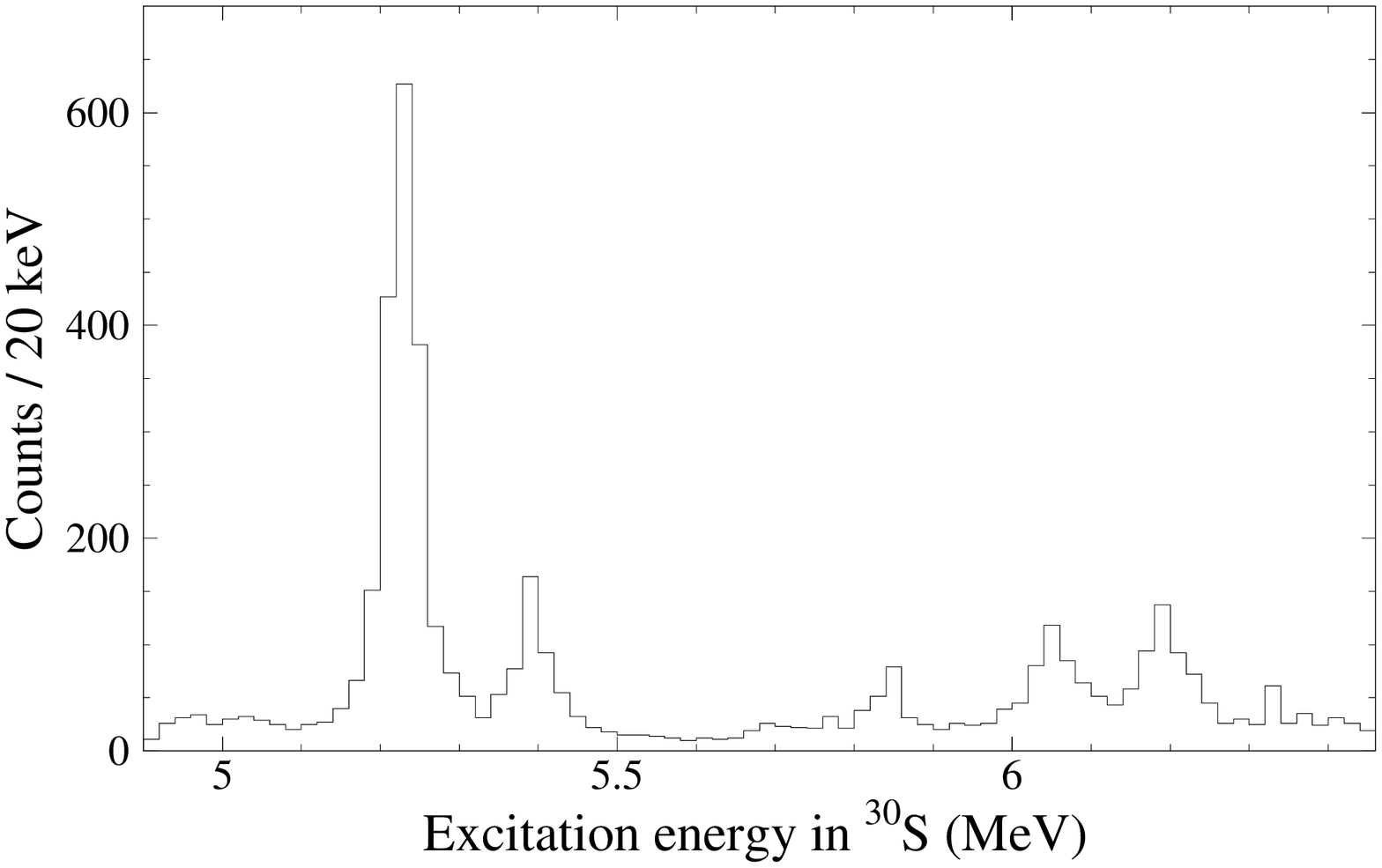} 
	\caption{Energy spectrum for \S calculated for the events from \fref{fynbofig}.}%
	\label{s30fig}%
\end{figure} 

\begin{figure}%
	\centering
	\includegraphics[width=0.90\columnwidth]{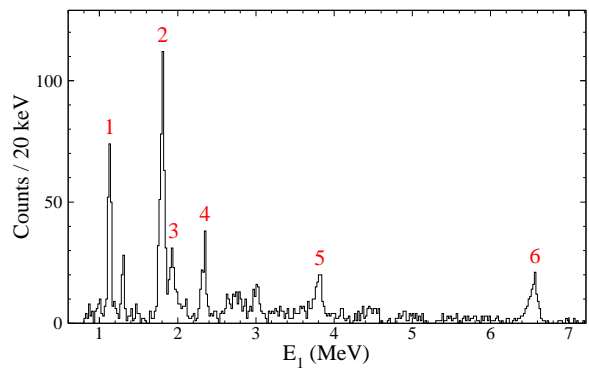} 
	\caption{(Colour online) Energy of the first particle for transitions going through the $\SI{5.2}{MeV}$-level in \S. The peaks containing most counts are marked by numbers.}%
	\label{e1fig}%
\end{figure}

\begin{figure}%
	\centering
	\includegraphics[width=0.90\columnwidth]{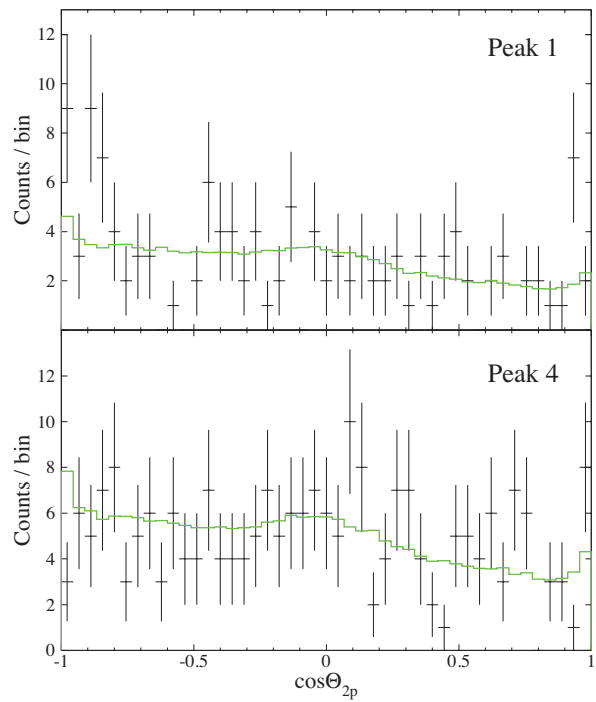} 
	\caption{(Colour online) Angular distribution of the two protons forming peak 1 and 4 in \fref{e1fig} compared with the corresponding uniform 2p-simulation. For better visualisation they are here shown using 45 bins, while the uniform fits are made using 90 bins.}%
	\label{ppfig}%
\end{figure} 

\begin{table}[bthp]
\caption{The $A_2$ coefficients for different 2p-transitions from \Cl
  trough the $\SI{5.2}{MeV}$-level in \S together with the difference
  in $\chi^2$ compared to a uniform fit and the result D of
  a Kolmogorov test to a uniform distribution. The peak
  numbers correspond to \fref{e1fig}.}
\begin{tabular}{c c c c c}
\toprule[0.1em]
\vspace{-3mm} \\
\vspace{0.8mm}Peak	&	E(\Cl) (MeV) 		& $A_{2}$		 &	$\Delta\chi^2$ & D		\\
\midrule[0.08em]
\vspace{-3.5mm} \\
1		& $6.674(6)$		&  $-0.12(14)$		& $0.67$		& $0.79$		\\
2		& $7.380(6)$		&  $0.16(11)$		& $1.97$		& $1.57$		\\
3		& $7.512(7)$		&  $0.35(19)$		& $3.51$		& $0.88$		\\
4		& $7.919(8)$		&  $0.48(19)$		& $6.69$		& $1.40$		\\
5		& $9.434(9)$		&  $0.04(19)$		& $0.05$		& $0.72$		\\
6 (IAS) & $12.313(4)$	&  $0.03(18)$		& $0.03$		& $0.65$		\\
\vspace{0.4mm}All &		&  $0.18(5)$		& $13.18$		& $7.95$		\\
\bottomrule[0.1em]
\end{tabular}
\label{pptab}
\end{table}


To use this method transitions from distinct levels in \Cl must be
identified with sufficient statistics. This is only possible for the
strongest fed level in \S at $\SI{5.2}{MeV}$. This level
has previously been assigned $0^+$ \cite{Seto2013}, but has
also been identified as a $3^+$ state due to its gamma decay
\cite{Lotay2012}. To have sufficient statistics all detectors are used
with a lower energy gate on the first particle of $\SI{800}{keV}$ and
the second of $\SI{500}{keV}$, except for the thick detector 5, where
$\SI{800}{keV}$ is used. The data can be seen in \fref{fynbofig}. The
\S levels calculated from these events can be seen in
\fref{s30fig}. The energy of the first particle (the one with the
highest energy) of the events passing through the $\SI{5.2}{MeV}$ level are
shown in \fref{e1fig}. The most intense peaks are numbered and are
used in the following analysis. In \fref{ppfig} the angular correlation for
two of the peaks are shown together with a simulation of the same decay
that assumes a uniform angular distribution (i.e. $A_2=0$). The simulated curves are fitted to
the data for all numbered peaks of \fref{e1fig} with and without
an $A_2$ term. The resulting $A_2$ values are shown in \tref{pptab}
along with the difference in $\chi^2$ for the two fits. Also shown are
the results of a Kolmogorov test (essentially the maximum difference in
cumulative distributions scaled with the square root of the number of
counts, the $\SI{5}{\%}$ significance level then corresponds to a value of $\num{1.36}$
\cite{gofbook}) for a comparison between the data and a
uniform distribution. Both the $\chi^2$ difference and the Kolmogorov
test indicate that the events in peak 1, 5 and 6 are
consistent with being uniform. The situation for peaks 2 and 3 is less
clear: The Kolmogorov test shows that the
events in peak 2 are not consistent with a uniform distribution with
$\SI{97.5}{\%}$ confidence \cite{gofbook}, but the deviations do not
correspond to a standard angular correlation shape since the fit does not give
a value for $A_2$ that are significantly different from $0$ (fits
including an $A_4$ term does not improve this).
The fit for the events of peak 3 points to a $A_2$
parameter different from $0$, but the Kolmogorov test does not
find the distribution to be significantly different from
uniform. Finally the events of peak 4 have a distribution significantly
different from uniform with more than $\SI{95}{\%}$ confidence
\cite{gofbook} using the Kolmogorov test and the value for $A_2$
is different from $0$ with more than $2\sigma$. This is also the case
if all events of the $\SI{5.2}{MeV}$ peak are considered, which
implies that there must be components that are non-uniform excluding the
$0^+$ and $1^+$ as a possibility for the spin of the \S-level by
comparison with \tref{A2tab}. Spin $2^+$ is also excluded since it can
only give deviations to negative $A_2$ values.

The spin of the states in \Cl is only known for the IAS where it is
$\frac{5}{2}^+$, the others can be either $\frac{3}{2}^+$,
$\frac{5}{2}^+$ or $\frac{7}{2}^+$ if one assumes that only allowed $\beta$-decays are observed experimentally. 
The $A_2$
value for the IAS (peak 6) indicates a uniform distribution. Comparing the value with \tref{A2tab} it is seen that the level in \S is either $0^+$, $1^+$, $2^+$ or $3^+$, but the $4^+$ cannot be completely
excluded due to the uncertainty on $A_2$. 
Comparing the value of $A_2$ for
peak 4 with \tref{A2tab} we can only conclude that the events stem from a $\frac{3}{2}^+$ or $\frac{5}{2}^+$ level in \Cl. If peak 4 corresponds to a
$\frac{5}{2}^+$ level the spin of the \S-level is $4^+$, and if it is
a $\frac{3}{2}^+$ level it is most likely a $3^+$ even though a $4^+$
cannot be completely excluded. 
Considering all the data in
\tref{pptab} and comparing them with \tref{A2tab}, the preferred value for the spin of the $\SI{5.2}{MeV}$ level is $3^+$, since we observe several levels that give a
uniform distribution, which for the $3^+$ level would be the case for
all decays from $\frac{5}{2}^+$ and $\frac{7}{2}^+$-levels in
\Cl. Furthermore the $A_2$ values that differ from $0$ are not too
high which should be the case for at least some of the decays through
a $4^+$ level due to decays from $\frac{3}{2}^+$-levels in
\Cl. Assuming that the $\SI{5.2}{MeV}$-level in \S is a $3^+$-level
implies that the $\SIE{7.919}{8}{MeV}$ level in \Cl is a
$\frac{3}{2}^+$ level and that the $\SIE{6.674}{6}{MeV}$ level is
either $\frac{5}{2}^+$ or $\frac{7}{2}^+$. The spin of the remaining three
levels (excluding the IAS) cannot be restricted due to the uncertainty
on the $A_2$ values.


\subsection{$\gamma$-transitions in the decay of \Artt}
\label{gamma33Ar}

\begin{figure}%
	\centering
	\includegraphics[width=0.90\columnwidth]{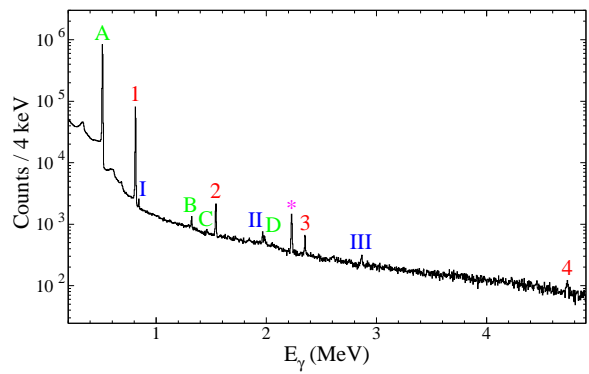} 
	\caption{(Colour online) The $\gamma$-spectrum of \Artt. The numbers corresponds to transitions in the $\beta$-daughter \Cltt, the Roman numbers to transitions in the $\beta$-granddaughter \ce{^{33}S} and the * to a transition in the $\beta$-proton daughter \Stt. The letters corresponds to background lines.}%
	\label{gam33fig}%
\end{figure} 

To obtain a good calibration of the particle detectors several runs with \Artt were made during the experiment. The $\gamma$-spectrum from these, i.e. in the decay of \Artt, can be seen in \fref{gam33fig}. The peak marked with numbers corresponds to transitions in the $\beta$-daughter \Cltt, the ones marked with Roman numbers to transitions in the $\beta$-granddaughter \ce{^{33}S} and the one marked by a * to a transition in the $\beta$-proton daughter \Stt. A, B, C and D are peaks from annihilation, pile-up and $\gamma$-transitions from decays of \ce{^{40}K} and \ce{^{18}N}, respectively. The assignment is supported by the half-life found for the peaks. The relative intensities of the $\gamma$-lines observed in the decay of \Artt are given in \tref{Ar33tab}. They are compared to results from three different experiments \cite{Adimi10,Borge87,Wilson80}. The interesting transition is the peak at $\SIE{4734.0}{20}{keV}$. This is the transition of the IAS in \Cltt to the first excited state in \Cltt, which has a $\gamma$-branching ratio of $\SIE{92}{8}{\%}$ \cite{Endt78}. This $\gamma$-transition has never been observed in the decay of \Artt, even though the branching ratio of \Artt to the IAS in \Cltt is $\SIE{31.0}{1}{\%}$ \cite{Adimi10}. 

\begin{table}[btp]
\caption{The relative branching ratios of the $\gamma$-transitions in the decay of \Artt (above the line) and \Cltt (below the line). The peak identifier corresponds to \fref{gam33fig}. The intensities of the $\gamma$-transitions from the \Artt decay are normalised to peak 1 and compared to the results of Ref. \cite{Adimi10}. The transition marked by a * is compared to Ref. \cite{Borge87} as suggested in Ref. \cite{Adimi10}. The intensities of the $\gamma$-transitions from the decay of \Cltt are normalised to peak III and compared to the results of Ref. \cite{Wilson80}.}
\begin{tabular}{c c c c c}
\toprule[0.1em]
\vspace{-3mm} \\
\vspace{0.8mm}Peak	&	$E_{\gamma}$ (keV) 		& $I_{\gamma}$		 	&	$E_{\gamma}^{\text{ref}}$ (keV) 		& $I_{\gamma}^{\text{ref}}$			\\
\midrule[0.08em]
\vspace{-3.5mm} \\
1		& $811.2(10)$		&  $100(10)$		& $810.6(2)$		& $100(1)$		\\
2		& $1541.0(10)$		&  $3.2(3)$			& $1541.4(6)$		& $3.6(2)$		\\
3		& $2342.3(11)$		&  $1.10(13)$		& $2352.5(6)$		& $1.3(2)$		\\
4		& $4734(3)$			&  $0.46(9)$			& 							&						\\
\vspace{0.4mm}*		& $2230.4(19)$		&  $3.9(4)$		& $2230.6(9)$		& $1.7(5)$		\\
\midrule[0.08em]
\vspace{-3.mm} \\
I		& $841.3(10)$		&  $109(16)$		& $841$		& $118.6(36)$		\\
II		& $1966.9(12)$		&  $132(18)$		& $1966$		& $104.2(16)$		\\
\vspace{0.4mm}III 	&  $2867(3)$		& $100(15)$		& $2866$				& $100.0(18)$	\\
\bottomrule[0.1em]
\end{tabular}
\label{Ar33tab}
\end{table}

\section{Summary}
\label{summary}
An improved half-life of \Ar has been found to be $\SIE{15.1}{3}{ms}$.

For the first time a spectroscopic analysis of the $\beta$-delayed three-proton decay of \Ar has been presented, showing that roughly half of the 3p-decays stem from the IAS in \Cl, while the rest stem from higher lying levels. It is shown that the decay is mainly sequential through the lowest levels above the proton threshold in \P, but a simultaneous component cannot be excluded.

A quantitative analysis of the $\beta$2p$\gamma$-decay has been performed, to search for $\gamma$-transitions from excited levels in \P. Only $\gamma$-rays from the two lowest excited levels in \P can be clearly identified, but there are no contradictions between what is seen in the 1p- and 2p-gated $\gamma$-spectra for the levels above them.

The experimentally measured Fermi strength using the decay of the IAS is improved including the $\beta$3p-decay and contributions from decays to higher-lying states in \S than previously observed. The total measured branching ratio in the decay is $\SIE{3.60}{44}{\%}$, which is lower than the theoretical value of $\SIE{4.24}{43}{\%}$, but they agree within one standard deviation. This leaves room for contributions from decays to excited states in \P (above $\SI{1.96}{MeV}$) and for a possible $\gamma$-decay of the IAS in \Cl.

A new method to determine the spin of low-lying levels in \S is presented. It uses angular correlations between the two protons in the $\beta$2p-decay passing through the level in question. Since the spin of the populated levels in \Cl is not known an ensemble of states in \Cl is used. The method is used for the level at $\SI{5.2}{MeV}$, which is found to be either a $3^+$ or $4^+$ level, with the data favouring the $3^+$. In previous studies it has been suggested to be a $0^+$ level \cite{Seto2013} from comparisons with the mirror nucleus, but it was identified as a $3^+$ level from its $\gamma$-decay \cite{Lotay2012}. It is currently not known if there might be two levels around this energy, but we can conclude that a $3^+$ level at $\SIE{5.227}{3}{MeV}$ is populated in the decay of \Ar. We see no indications that this peak may consist of two separate contributions.

Finally the $\gamma$-transitions in the decay of \Artt are measured and their relative intensities are given and compared to previous measurements. A new $\gamma$-line is found at $\SIE{4734}{3}{keV}$, which comes from the decay of the IAS and has not previously been identified in the decay of \Artt.

\section{Acknowledgement}
We thank Marek Pf\"utzner for helpful discussion and input on the analysis of the $\beta$3p-decay of \Ar.

This work was supported by the European Union Seventh Framework through ENSAR (Contract No. 262010). This work was partly supported by the Spanish Funding Agency under Projects No. FPA2009-07387, No. FPA2010-17142, and No. AIC-D-2011-0684, by the French ANR (Contract No. ANR-06-BLAN-0320), and by R{\'e}gion Aquitaine. A.S. acknowledges support from the Jenny and Antti Wihuri Foundation.

\flushleft 
\bibliography{ar31}

\end{document}